\newcommand{\ts}{TrainShield\xspace}

\documentclass[sigconf]{acmart}
\AtBeginDocument{%
  }

\usepackage{listings}

\usepackage{fontawesome5}
\usepackage{multirow, multicol}
\usepackage{xspace}
\usepackage{booktabs}

\usepackage{framed}
\usepackage{tabularx}
\usepackage{subcaption}
\usepackage[normalem]{ulem}

\copyrightyear{2026}
\acmYear{2026}
\setcopyright{cc}
\setcctype{by}
\acmConference[HT '26]{37th ACM Conference on Hypertext}{September 14--18, 2026}{London, United Kingdom}
\acmBooktitle{37th ACM Conference on Hypertext (HT '26), September 14--18, 2026, London, United Kingdom}
\acmDOI{10.1145/3800935.3830870}
\acmISBN{979-8-4007-2564-7/2026/09}
\acmConference[HT 2026]{37th ACM Conference on Hypertext and
Social Media}{September 14--18,
  2026}{London, England}

\begin{document}

\title{\ts: Targeted Awareness for Cybersecurity Training}

\author{Giovanni Pizzenti}
\authornote{This work was performed while the author was affiliated with Ermes Cybersecurity.}
\email{g.pizzenti@reply.it}
\affiliation{%
  \institution{Reply Spike}
  \city{Torino}
  \country{Italia}
}
\author{Alberto Verna}
\email{alberto.verna@polito.it}
\affiliation{%
  \institution{Politecnico di Torino}
  \city{Torino}
  \country{Italia}
}
\author{Nikhil Jha}
\email{nikhil.jha@polito.it}
\affiliation{%
  \institution{Politecnico di Torino}
  \city{Torino}
  \country{Italia}
}
\author{Giuseppe Tipaldo}
\email{giuseppe.tipaldo@polito.it}
\affiliation{%
  \institution{Politecnico di Torino}
  \city{Torino}
  \country{Italia}
}
\author{Stefano Traverso}
\email{s.traverso@ermes.company}
\affiliation{%
  \institution{Ermes Cybersecurity}
  \city{Torino}
  \country{Italia}
}
\author{Marco Mellia}
\email{marco.mellia@polito.it}
\affiliation{%
  \institution{Politecnico di Torino}
  \city{Torino}
  \country{Italia}
}

\renewcommand{\shortauthors}{Pizzenti et al.}

\begin{abstract}
In recent years, cybersecurity threats have increasingly exploited human behaviour rather than purely technical vulnerabilities, exposing the limits of traditional awareness programmes delivered outside real-world contexts. To bridge this gap, we introduce \ts, an interaction paradigm for contextual cybersecurity training that embeds adaptive learning interventions directly within user workflows. The system integrates real-time risk detection (e.g., phishing and data loss prevention) with event-triggered hypermedia overlays that dynamically connect users to context-specific learning nodes embedded within their browsing workflow to deliver personalised micro-learning content and structured feedback tailored to the user’s knowledge level and current context. This approach operationalises behavioural theories by transforming security incidents into immediate learning opportunities, shifting users from automatic to reflective decision-making at critical moments. We further formalise a design model that maps detected events to adaptive training instances, combining user modelling, context extraction, and large language model (LLM)-based content generation.

A preliminary study indicates that the approach is perceived as useful in increasing risk awareness and is preferred over lengthy and asynchronous traditional training formats, while also highlighting challenges in aligning generated content with user expectations. Overall, the results suggest that embedding contextual, event-driven training within everyday interactions is a promising direction for behaviour-oriented cybersecurity education.
\end{abstract}




\maketitle

\section{Introduction}


Cybersecurity is a critical concern for organisations and society at large. The scale of the threat is reflected in its economic impact: in 2024, the global average cost of a data breach reached 4.88 million USD, marking the largest increase since the pandemic \cite{ibm2024breach}. At the same time, attackers are evolving their capabilities, with 16\% of breaches involving AI-driven techniques such as automated phishing and deepfake impersonation \cite{ibm2025breach}.


Reducing this escalating challenge to a predominantly technical issue impedes progress: the failure to capture the multidimensionality of the phenomenon risks ``reinforcing the predominantly technical view of cybersecurity while separating disciplines that should be acting in concert to resolve complex cybersecurity challenges'' \cite[p.~18]{craigen2014defining}. The nature of cyberattacks is not a pure technical issue: while defence mechanisms improve, malicious actors are increasingly bypassing technical perimeters to directly target users, e.g., via increasingly complex phishing approaches. In this context, attackers do not merely exploit individual knowledge gaps, but strategically embed themselves in the everyday, technology-mediated routines of people. 
Attackers seek to insert themselves into the socio-technical routines through which people work, communicate, and make decisions, turning ordinary practices into potential vectors of compromise. In this context, \textit{security awareness} becomes the most important line of defence, where training is central. However, traditional training approaches remain largely decoupled from the contexts in which security decisions are made, limiting their effectiveness in shaping real-world behaviour.
What is missing is a model for integrating cybersecurity training directly into the interaction context in which security decisions occur.

Here, we propose \ts, an interaction model designed for event-triggered cybersecurity training, where content is adaptive and offers in-situ interventions embedded within user workflows. 
Unlike traditional awareness programmes
, \ts delivers training at the exact moment when a risky action occurs, transforming security incidents into immediate learning opportunities, in line with recommendations from the scientific literature \cite{prummer2024systematic, zhang2021systematic, lain2024content}.
\ts offers AI-generated short interactive hypermedia nodes, e.g., quizzes and micro-lessons, that are tailored to the security event and the user's skill level
. These interventions can be seen as dynamic, context-triggered nodes embedded within user workflows, where navigation is driven by user actions rather than predefined links.
Unlike traditional hypertext, where navigation follows explicit user-selected links, TrainShield creates implicit navigation paths triggered by user behaviour and security events, turning browsing actions themselves into the mechanism that connects users to educational content.

We conceptualise these interventions as contextual hypermedia overlays embedded in user workflows. While prior work explored phishing simulations and embedded training, these approaches remain episodic or pre-scripted, and do not adapt to real-time user context.
Our approach operationalises behavioural theories by introducing controlled interruptions that shift users from automatic to reflective decision-making at critical interaction points, working both as a defensive tool and as an educational resource. It helps users \textit{recognise} threats as they happen and allows them to \textit{learn} from these situations in a structured and engaging way. The main objective is to encourage the growth of a security-aware culture, where users are seen as a new line of defence against cyber threats rather than a weakness. This paper makes the following contributions:
\begin{itemize}
    \item We introduce an interaction paradigm for embedding cybersecurity training within user workflows through event-triggered interventions.
    \item We propose a design model mapping risk events to adaptive training content via user modelling and LLM-based generation.
    \item We provide an exploratory evaluation of perceived effectiveness and content quality.
    \item {We run a comparison among different LLM models, to assess the impact that the model choice has on the quality of the process.}
\end{itemize}

The rest of the paper is organised as follows: Section~\ref{sec:related} presents a review of background and related work from a socio-technical perspective. Section~\ref{sec:method} describes the rationale and the methodology behind the build-up of \ts, while Section~\ref{sec:result} discusses the most relevant results. Finally, Section~\ref{sec:conclusion} draws the final conclusions on the work.
\section{\textbf{A Socio-Technical Perspective on Security Awareness}}
\label{sec:related}

We conceptualise cybercrime as a socio-technical phenomenon emerging from interactions between users and digital systems \cite{gonzalez2015interdisciplinary, mouton2016social, nowak2022cybersecurity}.
From a hypertext perspective, cybersecurity incidents create opportunities for adaptive navigation, where user actions dynamically determine the educational resources presented. TrainShield operationalises this idea by coupling interaction events with context-aware hypermedia overlays. In this view, attacks are not purely technical exploits, but forms of communication that manipulate users within their everyday workflows.

In contemporary cybersecurity, many of the most pervasive threats target the human element through structured socio-technical manipulation \cite{hatfield2018social}. These strategies leverage socio-cognitive mechanisms such as trust, authority, and reciprocity \cite[p.~188]{mouton2016social}, embedding themselves within existing organisational practices. As a result, attackers exploit legitimate user behaviour rather than bypassing it, rendering purely technical countermeasures insufficient. As noted by Krombholz et al.~\cite[p.~114]{krombholz2015advanced}, social engineering operates by inducing users to disclose information or perform actions on behalf of the attacker.

\subsection{Phishing}

Social engineering refers to a class of attacks that exploit human behaviour through deceptive communication
. These attacks operate across multiple channels (e.g., email
, web interfaces, and more) and manipulate interactions within everyday digital environments \cite{krombholz2015advanced, mouton2016social}. Among these, phishing is one of the most widespread forms, where attackers impersonate trusted entities to induce users to disclose sensitive information or perform unsafe actions \cite[p.~117]{krombholz2015advanced}. While campaigns range from generic to highly targeted, they consistently exploit users’ expectations and routines in familiar contexts.

Their effectiveness is therefore primarily social rather than technical. By mimicking organisational communication patterns and leveraging cognitive biases such as trust and authority, attackers embed themselves within legitimate workflows. As a result, security failures often arise from routine user actions, highlighting the need for approaches that address behaviour in context, where user awareness remains a critical line of defence.

\subsection{Data Loss Prevention}
\label{sec:DLP}
Data Loss Prevention (DLP) encompasses technical and organisational mechanisms designed to prevent the unauthorised disclosure of sensitive information \cite{nistdlp}. Typical systems monitor user interactions and detect risky actions, such as sharing confidential data through messaging platforms or external services, including AI chatbots. DLP solutions rely on pattern matching and, in more advanced cases, context-aware analysis to identify potential data leaks. While effective at enforcing security policies, these systems primarily operate as blocking mechanisms, interrupting actions without necessarily improving user understanding.

As a result, DLP alone does not address the behavioural dimension of data exposure: users may remain unaware of why an action is risky or how to avoid similar situations in the future. This limitation highlights the need to complement detection and enforcement with mechanisms that provide immediate, context-aware feedback and training.
In this work, we build on this limitation by integrating DLP detection with contextual, in-situ training, transforming blocked actions into learning opportunities.

\subsection{Cybersecurity Awareness}

Cybersecurity \textit{awareness} refers to the dissemination of knowledge and practices aimed at helping individuals recognise threats and act appropriately, typically through broad communication strategies such as campaigns or informational materials \cite{nistsp80050}. In contrast, \textit{training} focuses on developing specific skills through structured activities, such as phishing simulations. Existing approaches span a range of formats, including games, simulations, and interactive content \cite{prummer2024systematic}, but their impact is often limited by a lack of contextualisation, as training is typically delivered \textit{outside} the situations in which security decisions are made.

Recent work on interactive formats, such as game-based training, shows improvements in user attention to security cues \cite{kavrestad2022evaluation}, highlighting the role of engagement in shaping behaviour. However, these approaches remain largely detached from real-world workflows, limiting their influence at the moment of risk. In this work, we address this limitation by integrating training directly into user interactions, combining adaptive content generation with real-time detection of risky events, and building on established instructional principles for cybersecurity education \cite{zhang2021systematic}.

Existing approaches can be grouped into four categories: traditional awareness programmes, phishing simulations, browser warnings, and security nudges. TrainShield differs by combining real-time detection, adaptive LLM-generated content, and embedded hypermedia interventions within a single interaction loop.

\subsection{Behavioural Theories and Cybersecurity Habits}
\label{sec:behavioural-theories}

Dual-process behavioural theory distinguishes between two complementary cognitive systems: the fast, automatic \emph{System 1} and the slower, analytical \emph{System 2}. In everyday digital interactions, most user actions are governed by System 1, relying on habits and heuristics rather than deliberate reasoning \cite{sharma2021impact}. 
Social engineering attacks exploit this tendency by inducing urgency or curiosity, keeping users in an automatic mode and reducing their ability to detect subtle security cues such as anomalous sender addresses or suspicious links. As a result, many security failures occur due to the context in which decisions are made.

This observation is particularly relevant for our setting, where security decisions are embedded within routine user workflows. To address this, \ts introduces brief, interaction-level interruptions (e.g., ``pause'' prompts) that require explicit user input, momentarily disrupting automatic behaviour and triggering reflective evaluation. 
The system complements these interventions with short, embedded training that provides immediate, context-specific feedback. This design aligns with prior findings on the effectiveness of in-situ training and behavioural nudges \cite{lain2024content}, while aiming to balance intervention strength with usability constraints to avoid cognitive overload.

\subsection{Just-in-Time Adaptive Interventions (JITAI)}

Our approach relates to recent work on context-aware and just-in-time learning systems, where educational interventions are triggered by the user’s current activity and context rather than delivered in isolated training sessions. In particular, Just-in-Time Adaptive Interventions (JITAI) provide a framework for delivering targeted support at critical decision points to influence user behaviour~\cite{hardeman2019systematic}. Similarly, research on smart and context-aware learning environments highlights how educational content can be dynamically adapted based on user state and situational factors \cite{hwang2014definition}. 

Unlike traditional JITAI systems, which typically operate within predefined domains (e.g., health or education), our approach applies these principles to security-critical interactions and couples them with real-time risk detection. Taken together, these perspectives highlight a common limitation: existing approaches either detect risks without fostering understanding, or provide training detached from the context in which decisions occur. This gap motivates the need for systems that integrate detection, context, and adaptive training within the same interaction loop.
\section{Methodology}
\label{sec:method}
\begin{figure}
    \centering
    \includegraphics[width=\columnwidth]{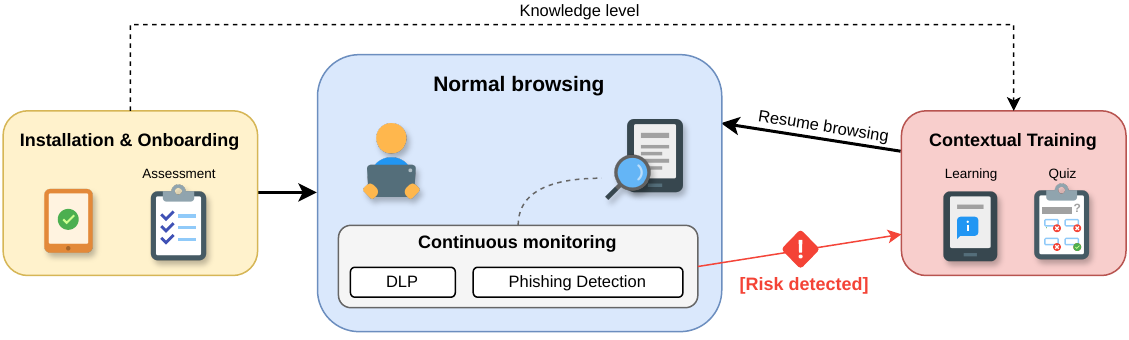}
    \caption{\ts process overview.}
    \label{fig:training-process}
\end{figure}

In this section, we present the methodology underlying \ts, focusing on how cybersecurity training is embedded within users' everyday web interactions. 
Our methodology follows a design-oriented approach, where we operationalise context-aware cybersecurity training through a pipeline that integrates (i) user modelling, (ii) real-time risk detection, and (iii) adaptive intervention. The goal is to study how event-triggered training can be systematically embedded into user workflows. We implement a proof-of-concept browser extension that enables real-time interaction with users. The system monitors browsing activity, identifies potentially risky situations (e.g., visiting suspicious web pages or attempting to share sensitive information) and converts these events into real-time learning opportunities. Figure~\ref{fig:training-process} illustrates the overall \ts workflow, describing how cybersecurity training is integrated into the user's browsing activity.
%
%
%
%
Formally, TrainShield can be modelled as a pipeline $P = \{U, D, C\}$, where $U$ is the user model (knowledge level), $D$ is the detection layer (event generation), and $C$ is the contextual training function mapping events to interventions.

Upon installation, the process begins with an {onboarding} phase, during which the user $U$ completes an initial assessment to estimate their level of cybersecurity knowledge (Section~\ref{sec:onboarding}).
After onboarding, 
the system operates transparently in the background, continuously monitoring interactions through multiple detection mechanisms $D$ (Section~\ref{sec:monitoring}). Our approach focuses on two mechanisms, i.e., phishing detection and data loss prevention (DLP), but can be extended to cover other classes of risk.
When a potentially risky situation is identified, the system triggers a {contextual awareness} session $C$, resulting in a temporary interruption of user activity during which the user is offered dynamically-generated training content (Section~\ref{sec:training}).
Over time, this cycle of monitoring, detection and contextual training allows for repeated exposure to real-world scenarios, supporting the gradual development of cybersecurity awareness.

\subsection{User Onboarding and Profiling}
\label{sec:onboarding}

The onboarding process combines a self-assessment questionnaire with a short quiz to estimate the user’s cybersecurity knowledge. The self-assessment captures perceived competence, while the quiz provides an objective measure, allowing the system to account for potential biases in self-evaluation \cite{podsakoff2003common}.



Users rate their familiarity (1–5) across three areas: Internet fundamentals, browsing, and phishing. Each area is then tested with one multiple-choice question. The final score is computed by weighting correct and incorrect answers by the corresponding self-assessment value, approximating a confidence-weighted knowledge estimate.

We address different but connected aspects: basic prerequisites (system), the environment (browser) and practical risks that come from its usage (phishing). The resulting values are saved and later used to compute the user level.
This approach allows us to take into account the user's \textit{confidence} in the three topics. Notably, if a user answers incorrectly to a question related to a topic they feel competent in, it highlights a case of overconfidence. Indeed, overestimation of one's own competence is not incidental but systematic, and disproportionately affects lower-skilled individuals \cite{kruger1999unskilled}. The scoring mechanism accounts for this by penalising misplaced confidence. The resulting scoring mechanism approximates a confidence-weighted knowledge estimate, where self-assessment acts as a proxy for perceived competence and is modulated by objective correctness.

\subsection{Continuous Monitoring and Risk Detection}
\label{sec:monitoring}

After onboarding, \ts operates during the user's normal browsing activity, continuously analysing interactions to potentially identify risky situations. This phase is designed to be transparent and non-intrusive, allowing the user to interact with web content without interruption unless a threat is detected.

The system relies on a modular detection layer composed of two modules: phishing detection and data loss prevention (DLP), two common and high-impact classes of security risk. The design is inherently extensible, allowing additional modules addressing other threat types to be integrated as needed.

\subsubsection{Phishing Detection}
\label{sec:phishing-detection}

The phishing detection module identifies potentially malicious web pages through a multi-factor analysis based on structural properties of the URL and page (see Table~\ref{tab:phishing-risk-factors} in Appendix~\ref{sec:appendix-phishing-detection}). Each feature contributes to an aggregate score with a weight reflecting its severity. The phishing score is computed as $S = \sum_i w_i f_i$, where  $f_i \in \{0,1\}$  indicates the presence of feature $i$ , and $w_i$ is its associated weight.

A page is classified as suspicious when the score $S\ge\theta$, $\theta$ being a predefined threshold, triggering a training event. In our methodology, detection is not treated as a classification task to be optimised, but as an event generation mechanism that activates contextual training. As such, false positives are acceptable, as they still produce learning opportunities.

\subsubsection{Data Loss Prevention}
\label{sec:dlp}

The DLP module focuses on protecting users from disclosing sensitive information during user interactions, primarily in text-based input fields. It operationalises the principles discussed in Sec.~\ref{sec:DLP}.
In this study, we focus on detecting sensitive information that is easily recognisable through pattern-based mechanisms (e.g., regular expressions), such as credit card numbers, IBANs and e-mail addresses. These checks are done in real time as the user interacts with input fields, allowing the system to identify potential data exposure before submission.








When sensitive content is detected on submission, the system blocks the associated action (e.g., disabling the submit button) and triggers contextual training. The action remains disabled until the sensitive portion of the message is deleted or the user acknowledges the correctness of the action.




\subsection{Contextual Training}
\label{sec:training}

This step defines a semantic abstraction layer that translates low-level detection signals into human-interpretable features suitable for instructional generation.
When a risky event is identified, \ts transitions from passive monitoring to active intervention by triggering a contextual training session. 
This step is performed directly within the browser environment, allowing the training to be delivered in the same context as the user's ongoing activity. This design is consistent with embedded simulation-based training paradigms (Section~\ref{sec:behavioural-theories}), where learning happens directly in the user's workflow.


\subsubsection{Threat Context Extraction}

Before generating the training content, the system builds a structured representation of the triggering event, which will be used to guide the generation process. For phishing events, the system translates the detected indicators (see Table~\ref{tab:phishing-risk-factors}) into human-readable explanations that capture the underlying context for the large language model (LLM). For instance, the ``Risky TLD'' feature is mapped to ``Uncommon top-level domain (.xyz, .top, ...)''. For DLP events, the context includes the type of sensitive information detected (e.g., credit card number, IBAN, etc.) and the nature of the attempted action.
In both cases, only the relevant features associated with the specific event are included, ensuring that the generated content remains focused and directly related to the user's behaviour.

\subsubsection{Training Content Generation}

The contextual information extracted from the previous step is used to construct a prompt to be sent to an LLM through a backend proxy (see Appendix~\ref{sec:llm-prompt-structure} for the full prompt structure). The prompt is structured to guide the model in producing consistent and context-sensitive outputs. To do so, we include background-level information such as:
(i) a description of the detected scenario $E$;
(ii) the user's knowledge level $U$ (i.e., \textit{Base} or \textit{Advanced});
(iii) the type of training objective $O$ (e.g., damage control, technological understanding, etc.), indicating what the question should focus on. The system rotates the question foci to avoid repetition in the quizzes, thus improving the long-term perceived usefulness of the system. We list all the foci in Appendix~\ref{app:question-focus};
(iv) constraints on format and content structure.

We thus define a mapping $C: (E, U, O) \rightarrow T$, where $E$ is the event context, $U$ the user model, $O$ the learning objective, $T$ the generated training instance.

This approach allows the system to generate varied training instances while maintaining coherence and alignment with the detected event. The use of a language model allows flexible generation of educational content in natural language, adapting to the specific scenario, user profile, and difficulty level. For instance, Base users are presented with simple text to understand, focusing on immediate actions and visible warning signs. 



\subsubsection{Adaptive Quiz and Feedback Mechanism}
\label{sec:quiz-feedback}



The quiz presents a multiple-choice question designed to prompt user reflection on the detected scenario. Each question includes four answers, of which only one is correct. 
%
%
The wrong answers are generated with the following possible flaws
: incomplete action, wrong sequence of actions, technical errors, wrong focus, excessive response, insufficient response, common yet incorrect belief.

The system also provides feedback organised into three components:
(i) \textbf{Why}: an explanation of why the event was blocked;
(ii) \textbf{Risks}: a concise description of potential consequences, giving a concrete damage estimation;
(iii) \textbf{Prevention}: a short list of solutions to avoid similar actions in the future, such as the use of tools or corrective behaviours.
This tripartite structure operationalises micro-learning principles by combining explanation, consequence framing, and actionable guidance within a constrained interaction window. It aims to reinforce understanding by clearly linking the user's action to its implications and immediate suggestions on how they can be prevented.

\subsubsection{Postponement and Override Mechanisms}

To support real-world usage, the system includes mechanisms that enhance user autonomy during training. The \textit{postponement} option allows users to defer a session when immediate completion is not feasible, storing the event for later presentation. For DLP events, an \textit{override} mechanism (``\textit{Send Anyway}'') enables users to proceed after completing the training and acknowledging the associated risks.

These features address both false positives and legitimate user needs, while reflecting findings from \cite{lain2024content} on the negative impact of excessive training pressure. Overall, they introduce a trade-off between enforcement and autonomy, which is critical for sustaining long-term engagement.

\subsection{User testing}\label{sec:user-testing}


To assess the effectiveness of TrainShield, we conduct an exploratory user study in which we evaluate the following hypotheses:
(H1) Contextual, in-situ training improves perceived risk awareness;
(H2) Event-triggered interventions are preferred over traditional training approaches;
(H3) LLM-generated content is sufficiently accurate and relevant for cybersecurity training.
For this, we rely on a human panel of users that we invited to install and test \ts over approximately two weeks. During this period, they encountered simulated or naturally occurring events triggering training interventions. At the end of the session, they completed an anonymous structured questionnaire.

Rather than performing a controlled comparison, we assess perceived effectiveness relative to traditional approaches through self-reported measures. We combine quantitative metrics (Likert scale responses) with qualitative feedback (open-ended suggestions) to capture both measurable performance indicators and more detailed user experience feedback. 
%
%
%
Opinions 
were summoned by means of a Google Form inspecting the extension's effectiveness, both from a technical and from an awareness-specific point of view. 
For all of 
the quantitative questions, the user can answer with a score from 1 to 5. At the survey end, an optional open text box allows the user to input additional information to better interpret the assigned score. These metrics are grouped into three dimensions:
(i) usability (U1–U2),
(ii) content effectiveness (E1–E5),
(iii) perceived impact (A1–A4).
Since the quality of contextual training depends on the underlying language model, we later compare several state-of-the-art LLMs to assess their suitability for this task.

The study was conducted in accordance with standard research ethics practices. No personal data was collected, and all participants provided informed consent for anonymous data collection. We involved a total of 18 testers, including both employees at a company and colleagues from the University's computer science department. Participants have heterogeneous backgrounds ranging from admin and non-technical roles to cybersecurity-aware profiles.

\begin{table}
    \centering
    \small
    \caption{User feedback survey results. Details about the metrics are in Appendix~\ref{app:questions}.}
    \resizebox{\columnwidth}{!}{%
    \begin{tabular}{clcc}
    \toprule
    
    & \textbf{Metric} & \textbf{Hypotheses} & \textbf{Avg. Score} \\
    \midrule
    \multirow{4}{*}{{\rotatebox[origin=c]{90}{\parbox[c]{4em}{\centering Usability \\ \& Design}}}}
        \\
        & (U1) Interface Clarity & - & 4.33 $\pm$ 0.69 \\
        & (U2) Technical Reliability & - & 3.83 $\pm$ 1.54 \\
        \\
    \midrule
    \multirow{5}{*}{{\rotatebox[origin=c]{90}{\parbox[t]{5em}{\centering Effectiveness}}}}
        & (E1) Onboarding Accuracy & - & 4.78 $\pm$ 0.43 \\
        & (E2) Quiz Relevance and Clarity & H3 & 3.78 $\pm$ 1.11 \\
        & (E3) Quiz Difficulty & H3 & 3.11 $\pm$ 0.68 \\
        & (E4) Educational Message Clarity & H3 & 3.72 $\pm$ 1.07 \\
        & (E5) Content Detail Appropr. & H3 & 4.39 $\pm$ 0.92 \\
    \midrule
    \multirow{4}{*}{{\rotatebox[origin=c]{90}{\parbox[t]{4em}{\centering Approach \\ Validation}}}}
        & (A1) Risk Perception & H1 & 4.22 $\pm$ 0.88 \\
        & (A2) Real-time Detection & H1 & 4.44 $\pm$ 0.78 \\
        & (A3) Context. v. Tradit. Training & H2 & 4.17 $\pm$ 0.99 \\
        & (A4) Learning Enhancement & H3 & 3.89 $\pm$ 1.18 \\
    \bottomrule
    \end{tabular}
    }
    \label{tab:extension-feedback-scores}
\end{table}

\section{Results}
\label{sec:result}


In this section, we present the result of the user testing we introduced in Section~\ref{sec:user-testing}, together with a a quantitative comparison between different LLMs to offer a first insight into the impact that the choice of the model could have on the generation of awareness quizzes
. At the time of testing, GPT-4o offered the best trade-off between costs and accuracy.

\subsection{Users' feedbacks on \ts}

Table~\ref{tab:extension-feedback-scores} presents the results of the survey, together with the hypothesis each question refers to. 
%
%
While appreciating the extension's UI, some participants presented minor technical problems: quiz questions reappearing after submitting the answer, data sanitisation failures, and other minor aspects that we fixed in a subsequent release of the extension.




Note that question E3, which focuses on quiz difficulty, has a scale ranging from 1 (way too easy) to 5 (way too difficult), with the median value 3 indicating an appropriate difficulty.
An average score of 3.11 shows an overall satisfaction. This is also mirrored by high scores given to the accuracy of the onboarding process (E1) in evaluating one's knowledge label and to the level of technical detail used by the extension in presenting information (E4, E5).

Overall, high scores in A1 (risk perception) and A2 (real-time blocking utility) support H1, suggesting that contextual interventions effectively increase users’ awareness during interaction. The section on approach validation (A1--A4) shows an appreciation of the contextual training proposal, validating H2. 

While positive, users were less enthusiastic about the relevance and clarity of some questions, and about the clarity of the educational messages provided after answering the quiz. In comments, participants assigning scores of 3 or below indicated that sometimes it was difficult to find a connection between the prevented action and the quiz topic, leading to distraction. Also, other users noticed that some quizzes were sometimes not suited for scenarios| where some actions (e.g., installing a VPN) may collide with company policies. Overall, lower scores in E2 and E4 indicate limitations in the alignment between detected events and the content generated by the LLM, highlighting a key challenge in context-sensitive generation (H3)---which we investigate in Section~\ref{sec:llm-comparison}.




Overall, results suggest that while the interaction paradigm is well-received, content generation quality remains the primary bottleneck. The approach is perceived as useful by the majority of users. A relatively low score about the effectiveness of the quiz---although satisfactory altogether---echoes the hurdles in creating a context-aware, informative quiz. In the next Section, we compare how different LLMs perform.

\subsection{LLM  comparison}
\label{sec:llm-comparison}

\begin{figure}
    \centering
    \includegraphics[width=\linewidth]{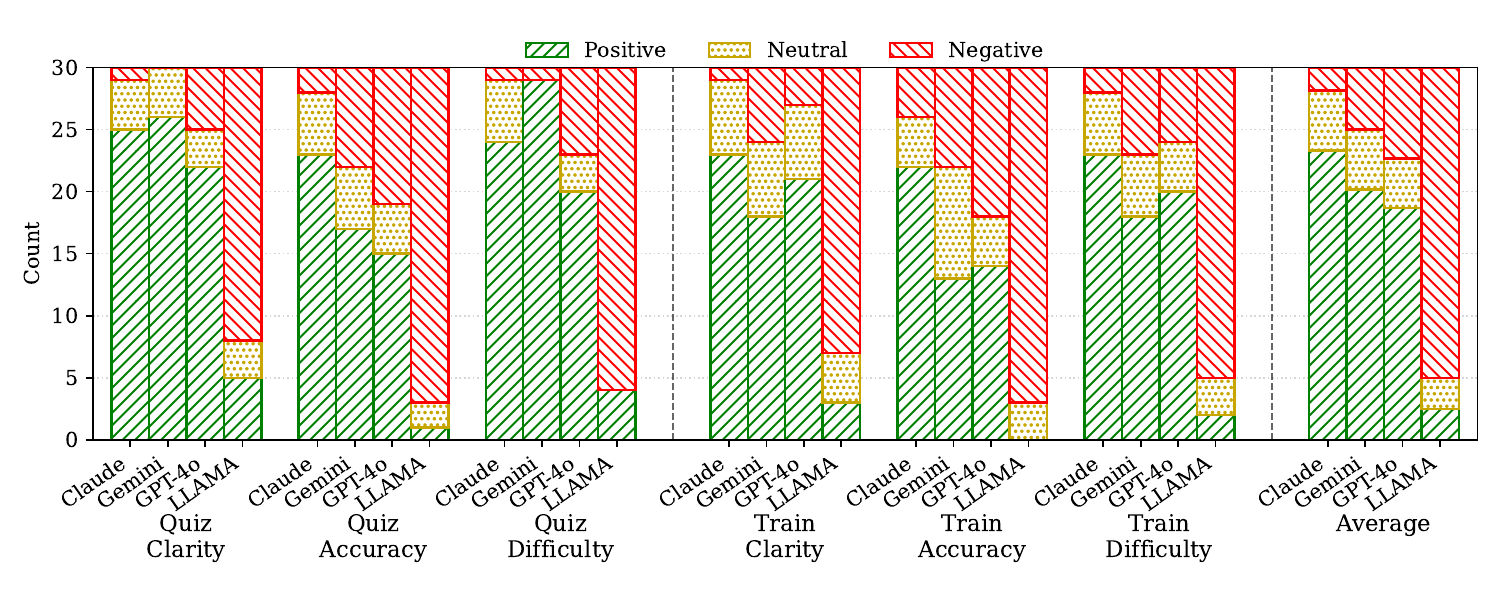}
    \caption{LLM model comparison.}
    \label{fig:model-comparison}
\end{figure}

This experiment isolates the impact of the content generation component within our methodology. We run a comparison among different LLM models to investigate the impact that the model choice has on the quality of the proposed training sessions. We use models from multiple vendors and with different sizes, including Claude Sonnet 4.6, Gemini 3 Flash, and LLAMA 3.1 8B Instruct, and compare them with GPT-4o that we used in our user-driven evaluation.

We isolated 10 separate quiz and training content prompts for various detected events and user expertise. We then ask each model to provide a quiz and training content (see details about the prompts in Appendix~\ref{sec:llm-prompt-structure}). We obtain 40 model outputs that we ask six cyberthreat intelligence experts to evaluate: each item is assigned to three reviewers to reduce individual bias. We make sure to rotate reviewer triplets, to minimise the number of items evaluated by the same subset of reviewers.

Each reviewer evaluates both the quiz and the training content for the assigned output, focusing on i) Clarity, ii) Accuracy, and iii) Difficulty. The Accuracy dimension refers both to the technical correctness of the output and to its relevance to the context. For instance, if the detected event is the visit to a page with a suspicious hyphenation and the quiz asks a question about DLP, the accuracy score will be low---irrespective of the actual correctness of the question and answers. Reviewers are asked to report for each dimension a value ranging from 1 to 5. We later map these values to negative (1--2), neutral (3), and positive (4--5) feedback.

Figure~\ref{fig:model-comparison} shows the result of our experiment. Average results (top-right breakdown) show that large models exhibit comparable performance, with Claude Sonnet 4.6 taking the lead. The small LLAMA 8B model fails to meet requirements. Looking at quiz category results, Gemini 3 Flash shows better figures than Claude Sonnet 4.6 in Clarity and Difficulty. GPT-4o---the best model at the time of running our initial experiments---is now outdated.

Moving to the Train questions, we see that generating high-quality training content is slightly more complicated. Here, Gemini has lower performance in both on Clarity and Difficulty with respect to quizzes. A double-check on the train sessions generated by Gemini 3 Flash reveals that they are in fact more generic and less specific than those generated by Claude Sonnet 4.6.

{We use Krippendorff's $\alpha$ to measure the agreement among reviewers. For all six dimensions, $\alpha$ is in the range 0.34--0.47 with an average of 0.39, showing a moderate agreement---yet, an above-random one. This is expected because reviewers evaluate pedagogical quality rather than objective correctness alone.}

\section{Limitations and Conclusions}
\label{sec:conclusion}
This work introduced \textit{TrainShield}, a context-aware cybersecurity training system that embeds adaptive learning interventions within user workflows. By coupling real-time risk detection with dynamically generated content, the approach reframes awareness as a continuous, in-situ learning process. Results from a preliminary user study suggest that this paradigm is perceived as useful, particularly in improving risk awareness and favouring contextual training over traditional approaches. The LLM comparison further highlights the importance of model capacity for generating accurate and relevant training content.

However, several limitations must be acknowledged. The evaluation is based on a small sample ($n = 18$) and relies on self-reported measures, limiting generalisability and preventing conclusions about actual behavioural change. No controlled baseline is included, and detection mechanisms are not optimised for accuracy, as they serve primarily to trigger training events. Additionally, the content generation pipeline shows limitations in aligning training with the triggering context.

Future work should address these limitations through controlled studies, improved alignment between detection and content generation, and more refined user modelling. Extending the system to incorporate organisational policies would further increase its applicability in real-world settings. Similarly, we currently rely on prompt engineering to constrain generated content. Future work will investigate automated validation and organisation-specific policy checking before content is delivered to users.

Overall, this work provides an initial step towards embedding adaptive cybersecurity training into everyday interactions, suggesting that contextual, event-driven learning is a promising direction for addressing the behavioural dimension of cybersecurity.

\begin{acks}
This work has received funding from the Applied Sciences Italian Fund (Fondo Italiano per le Scienze Applicate---FISA) by the Italian Ministry of University and Research, under the AI4CTI project (grant agreement No. FISA-2023-00168).
\end{acks}

\bibliographystyle{ACM-Reference-Format}
\bibliography{bibliography}

\appendix
\section{Phishing Detection Features}
\label{sec:appendix-phishing-detection}

In Table~\ref{tab:phishing-risk-factors}, we report the phishing detection figures and their scores. If a visited website overcomes a threshold $\theta$ ($\theta=8$ in our test), we classify it as phishing.

\begin{table*}
    \caption{Phishing detection features and associated weights used in the scoring mechanism.}
    \label{tab:phishing-risk-factors}
    \centering
    \begin{tabularx}{\textwidth}{crXc}
        \toprule
         & \textbf{Factor} & \textbf{Description} & \textbf{Weight} \\
        \midrule
        \multirow{19}{*}{{\rotatebox[origin=c]{90}{\centering URL \& Page}}}
            & \textbf{IP host} & The hostname is a raw IP address (e.g. \texttt{104.18.3.24}) instead of a domain & 3 \\
            & \textbf{Risky TLD} & Unusual top-level domain (e.g. \texttt{.xyz}) & 3 \\
            & \textbf{Brand mismatch} & A known brand name appears in subdomains or labels instead of the second level (e.g., \texttt{paypal.security-login.com} instead of \texttt{paypal.com}) & 4 \\
            & \textbf{Typosquatting} & URL contains a minor change from a known brand name (e.g., \texttt{paypa1.com}) & 4 \\
            & \textbf{Deep subdomain chain} & The URL has more than 3 sub-domains (e.g. \texttt{secure.login.product.aliexpress.com}) & 1 \\
            & \textbf{High number / hyphen presence} & High amount of numbers / hyphens are present in the hostname (e.g. \texttt{1a2b3c456789.pages.dev}) & 1 \\
            & \textbf{Long hostname} & Host name contains more than 30 characters & 1 \\
            & \textbf{Suspicious keywords} & Terms commonly used in phishing pages are present in the URL (e.g., ``login'', ``secure'', ``verify'') & 2 \\
            & \textbf{Title-domain inconsistency} & The brand name that appears in the page title is not the same as that in the host name & 2 \\
            & \textbf{Password field} & The page contains a password field. & 2 \\
            & \textbf{Cross-domain navigation} & The loaded tab URL differs from the one previously visited & 2 \\
            \midrule
            
        \multirow{4}{*}{{\rotatebox[origin=c]{90}{\centering \parbox[c]{4.5em}{Domain \& \\ Certificate}}}}
            & \textbf{Young domain} & The domain is younger than 180 days (30 days) & 4 (6) \\
            & \textbf{Self-signed certificate} & The page's TLS certificate is self-signed & 3 \\
            & \textbf{Free certification authority} & The page's TLS certificate is issued by a free certification authority (e.g., Let's Encrypt) & 1 \\ 
        \bottomrule
    \end{tabularx}
\end{table*}

\section{LLM Prompt Structure}
\label{sec:llm-prompt-structure}

We generate training content and quizzes through a structured prompt template that combines fixed instructional components with contextual information such as the detected event, the user's expertise level and the selected learning objective.

The prompt is organised into multiple blocks, each designed to control a specific aspect of the generated output.

\subsection{Role and Expertise}

\begin{framed}
You are a cybersecurity expert creating educational content for \texttt{<EXPERTISE>}-level users.
\end{framed}

The \texttt{<EXPERTISE>} field represents the user's knowledge level, which in our current implementation can be either \textit{Base} or \textit{Advanced}.

\subsection{Scenario}

This prompt section provides context on the event that triggered the training event. Depending on the event, the content is set to one of the following:
\begin{itemize}
    \item ``The user submitted an email, which is not authorised on chatbot like ChatGPT.''
    \item ``The user is browsing a phishing page identified by an IP visible in the URL.''
    \item ``The user submitted a Credit Card Number, which is sensitive data, but is not authorised to do so.''
    \item ``The user is browsing a phishing page identified by a rare TLD (.xyz, .top).''
    \item ``The user submitted a Social Security Number, which is sensitive data, but is not authorised to do so.''
    \item ``The user is browsing a phishing page identified by a self-signed TLS certificate.''
    \item ``The user submitted an IBAN, which is sensitive data, but is not authorised to do so.''
    \item ``The user is browsing a phishing page identified by suspicious hyphens in the domain.''
    \item ``The user is browsing a phishing page identified by a long hostname.''
\end{itemize}

\subsection{Question Focus}
\label{app:question-focus}

This block defines the learning objectives and lessons that the question and training content should convey. This part is added to avoid the model generating the same questions, introducing variation across training instances. The content is randomly selected among the following:

\begin{itemize}
    \item ``Immediate safety and damage control.''
    \item ``Long-term security posture improvement.''
    \item ``Technical understanding and analysis.''
    \item ``User awareness and education.''
    \item ``Organisational policy and compliance.''
    \item ``Incident response and recovery.''
    \item ``Potential consequences of the attack.''
    \item ``Prevention strategies.''
    \item ``Best practices.''
\end{itemize}

\subsection{Answer Quality}

\begin{framed}
    \begin{itemize}
        \item \textit{one} clearly correct answer;
        \item \textit{three} wrong or plausible but flawed alternatives;
        \item avoid ambiguous or subjective choices.
    \end{itemize}
\end{framed}

\subsection{Question Complexity}

\begin{framed}
    For \textbf{\textit{Basic}} users (no cybersecurity background):
    \begin{itemize}
        \item Use \textit{simple, everyday language}, avoid jargon;  
        \item focus on \textit{immediate practical actions};
        \item ask about \textit{recognisable warning signs}.  
    \end{itemize}

    For \textbf{\textit{Advanced}} users (cybersecurity expert):
    \begin{itemize}
        \item use \textit{technical terminology} and industry concepts.
    \end{itemize}
\end{framed}

These instructions operationalise the \texttt{<EXPERTISE>} parameter by guiding the language and conceptual depth of the generated content.

\subsection{Training Content Guidelines}

\begin{framed}
\textbf{WHY} (1 sentence, $\leq$25 words)
\begin{itemize}
    \item \textit{specific attack mechanism} and \textbf{blocking reason} with technical detail.
\end{itemize}

\textbf{RISKS} (exactly 2 bullets, $\leq$15 words each)
\begin{itemize}
    \item \textit{concrete scenario} for the end user;
    \item \textit{quantifiable business impact} and \textit{technical compromise chain};
    \item Use at least \textit{1 bold technical term} per bullet.
\end{itemize}

\textbf{PREVENTION} (exactly 2 bullets, $\leq$15 words each)  
\begin{itemize}
    \item \textit{specific preventive actions} (tools/controls).; 
    \item \textit{predictive best practice} (no generic advice); 
    \item \textit{imperative, actionable} language;
    \item use at least \textit{1 bold technical term} per bullet.
\end{itemize}
\end{framed}

This structure mirrors the feedback mechanism described in Section~\ref{sec:quiz-feedback}, guiding the model in generating explanations that directly link the user's action to its consequences and mitigation strategies.

\subsection{Quiz Distractors}
\label{app:distractors}

\begin{framed}
Create wrong answers using these specific approaches:
\begin{enumerate}
    \item \textbf{Type A}: \textit{incomplete action} --- partially correct but missing critical steps.
    \item \textbf{Type B}: \textit{wrong sequence} --- correct actions in wrong order or timing.
    \item \textbf{Type C}: \textit{technical error} --- technically incorrect claims.
    \item \textbf{Type D}: \textit{wrong focus} --- potentially right action, that does not refer to the current threat.
    \item \textbf{Type E}: \textit{excessive response} --- excessive response that could cause other problems.
    \item \textbf{Type F}: \textit{insufficient response} --- action that does not suffices in tackling the problem.
    \item \textbf{Type G}: \textit{common yet incorrect belief} --- action based on common belief, that does not solve the problem.
\end{enumerate}
\end{framed}

This aspect is added to improve the questions' quality: instead of relying on the model to choose distractors, we assist it by providing three common failure modes in the cybersecurity decision-making process.

\subsection{Technical Constraints}

\begin{framed}
    \begin{itemize}
        \item Language: English.
        \item Use \textit{$\geq$3 bold technical terms} across risks + prevention.
    \end{itemize}
\end{framed}

\section{Evaluation questionnaire}
\label{app:questions}
Here we summarise the questions included in the evaluation questionnaire:

\begin{itemize}
    \item Usability and interface design:
    \begin{itemize}
      \item \textbf{(U1) Interface Clarity}: Measures the clarity and ease of use of the extension;
      \item \textbf{(U2) Technical Reliability}: Evaluates the absence of bugs, crashes, or unexpected behavior during usage;.
    \end{itemize}
    \item Product effectiveness:
    \begin{itemize}
      \item \textbf{(E1) Onboarding Accuracy}: Evaluates whether the initial phase correctly identified participants' actual knowledge levels;
      \item \textbf{(E2) Quiz Relevance and Clarity}: Evaluates whether AI-generated questions are appropriate and understandable;
      \item \textbf{(E3) Quiz Difficulty}: Assesses the appropriateness of quiz difficulty;
      \item \textbf{(E4) Educational Message Clarity}: Measures the quality and structure of training content provided after risky actions;
      \item \textbf{(E5) Content Detail Appropriateness}: Evaluates whether the level of technical detail matches participants' knowledge levels.
    \end{itemize}
    \item Approach validation and perceived usefulness:
    \begin{itemize}
      \item \textbf{(A1) Risk Perception}: Assesses whether \ts improves users' awareness of cybersecurity threats during browsing;
      \item \textbf{(A2) Real-time Blocking Utility}: Measures perceived value of blocking risky actions at the moment they occur;
      \item \textbf{(A3) Contextual vs Traditional Training}: Direct comparison between contextual, real-time training and traditional theoretical cybersecurity courses;
      \item \textbf{(A4) Quiz/Messages Learning Enhancement}: Evaluates whether personalized quiz content and educational messages improve learning outcomes.
    \end{itemize}
\end{itemize}

\end{document}